\documentclass[reprint,showpacs,preprintnumbers,amsmath,amssymb,a4paper,nofootinbib,superscriptaddress,floatfix,aps,pra,longbibliography,showkeys]{revtex4-1}

\setcitestyle{numbers,open={[},close={]}}

\usepackage{bm}
\usepackage{slashed}
\usepackage{amsfonts}
\usepackage{wasysym}
\usepackage{color}
\providecommand{\U}[1]{\protect\rule{.1in}{.1in}}
\providecommand{\U}[1]{\protect\rule{.1in}{.1in}}
\newcommand{\be}{\begin{equation}}
\newcommand{\en}{\end{equation}}

\usepackage{todonotes}
\usepackage{txfonts}	
\usepackage[colorlinks=true, pdfstartview=FitV, linkcolor=blue, citecolor=blue, urlcolor=blue]{hyperref}

\begin{document}

\preprint{$\today$}

\title{Tensile deformations of the magnetic chiral soliton lattice probed by Lorentz transmission electron microscopy}

\author{G. W. Paterson}
    \affiliation{SUPA, School of Physics and Astronomy, University of Glasgow, Glasgow, G12 8QQ, UK}
\author{A. A. Tereshchenko}
    \affiliation{Institute of Natural Science and Mathematics, Ural Federal University, Ekaterinburg, 620002, Russia}
\author{S. Nakayama}
    \affiliation{Department of Physics and Electronics, Osaka Prefecture University, 1-1 Gakuencho, Sakai, Osaka, 599-8531, Japan}
\author{Y. Kousaka}
    \affiliation{Department of Physics and Electronics, Osaka Prefecture University, 1-1 Gakuencho, Sakai, Osaka, 599-8531, Japan}
\author{J. Kishine}
    \altaffiliation{Institute for Molecular Science, 38 Nishigo-Naka, Myodaiji, Okazaki, 444-8585, Japan}
    \affiliation{Division of Natural and Environmental Sciences, The Open University of Japan, Chiba, 261-8586, Japan}
\author{S. McVitie}
    \affiliation{SUPA, School of Physics and Astronomy, University of Glasgow, Glasgow, G12 8QQ, UK}
\author{A. S. Ovchinnikov}
    \altaffiliation{Institute of Metal Physics, Ural Division, Russian Academy of Sciences, Ekaterinburg, 620219, Russia}
    \affiliation{Institute of Natural Science and Mathematics, Ural Federal University, Ekaterinburg, 620002, Russia}
\author{I. Proskurin}
    \altaffiliation{Institute of Natural Science and Mathematics, Ural Federal University, Ekaterinburg, 620002, Russia}
    \affiliation{Department of Physics and Astronomy, University of Manitoba, Winnipeg, Manitoba, R3T 2N2, Canada}
\author{Y. Togawa}
    \affiliation{Department of Physics and Electronics, Osaka Prefecture University, 1-1 Gakuencho, Sakai, Osaka, 599-8531, Japan}

\date{\today}

\begin{abstract}
We consider the case of a chiral soliton lattice subjected to uniaxial elastic strain applied perpendicular to the chiral axis and derive through analytical modelling the phase diagram of magnetic states supported in the presence of an external magnetic field.
The strain induced anisotropies give rise to three distinct non-trivial spin textures, depending on the nature of the strain, and we show how these states may be identified by their signatures in Lorentz transmission electron microscopy (TEM).
Experimental TEM measurements of the Fresnel contrast in a strained sample of the prototypical monoaxial chrial helimagnet CrNb$_3$S$_6$ are reported and compare well with the modelled contrast.
Our results demonstrate an additional degree of freedom that may be used to tailor the magnetic properties of helimagnets for fundamental research and applications in the areas of spintronics and the emerging field of strain manipulated spintronics.

\end{abstract}

\maketitle

\section{Introduction}
The connection between mechanical stress and magnetic subsystems in solids is at the core of a new branch of electronics, strain manipulated  electronics, sometimes referred to as straintronics \cite{Si2016,Bukharov2018}.
Through manipulation of the spin degrees of freedom, it is expected that strain engineering of the magnetic properties of spintronic devices will be able to realise ultralow energy consumption devices approaching the limit imposed by fundamental principles \cite{Landauer1961}.
One of the significant ideas of this research area is to control magnetization on the basis of the magnetoelastic effect by the formation of an additional magnetic anisotropy evoked by mechanical stresses \cite{Madami2017}.
The induced strain field provides an alternative to both the magnetic field and the spin transfer torque produced by a current.
This approach has been realized in a wide range of systems, including the switching dynamics of single-domain nanoparticles by stress \cite{Roy2011}, by ultrafast acoustic pulses \cite{Kovalenko2013}, and by tensile strain induced from a piezoelectric substrate \cite{Buzzi2013,Souza2016,Tiercelin2011}.
Strain engineering techniques have been successfully employed to manipulate magnetic moments in perovskite-based multiferroics \cite{Ramesh2007,Sando2013,Gareeva2013,Agbelele2017,Prellier2005,Kawachi2017}, multiferroic thin-film heterostructures \cite{Flebig2016}, and in nanowires \cite{Lopez2013}. 
A change in the magnetic anisotropy governed by strain is clearly demonstrated in ferromagnetic-ferroelectric \cite{Shirahata2013,Moubah2013} and ferrite-ferroelectric heterostructures \cite{Pan2013} due to structural or metal-insulator \cite{Venta2013} phase transformations in the material.
In addition, strain has long been of interest in superconductors, with recent work having shown it may be used to alter the pair formation mechanism \cite{Llordes2012} and the transition temperature \cite{Ahadi2019} of different materials.
Furthermore, a great variety of applications are related to the development of stretchable electronics, in which intrinsic flexible functional materials are able to operate under mechanical strain \cite{Rogers2010,Wu2019,Liu2014}. 

An effect of elastic stresses on the magnetic properties of chiral helimagnets is of growing interest. 
Recently, it was demonstrated that a skyrmion crystal (SkX) in a cubic chiral helimagnet is very sensitive to deformations of the underlying crystal used to stabilize the non-trivial spin texture.
Lorentz transmission electron microscopy (TEM) studies of the magnetic configuration revealed that SkX distortions are amplified by two orders of magnitude in comparison with elastic strains in the crystal lattice \cite{Shibata2015}.
This provides a new approach of skyrmion crystal manipulation by elastic lattice degrees of freedom  \cite{Fobes2017,Kang2017,Zhang2017,Kanazawa2016,Hu2017,Hu2019}.

Helimagnets of hexagonal symmetry, such as CrNb$_3$S$_6$, subjected to an external magnetic field exhibit another type of nontrivial magnetic order, the magnetic soliton lattice \cite{Dzyaloshinskii1964, Izyumov1984}.
Magnetostrictive deformations in the monoaxial chiral helimagnet were addressed in Refs.~\onlinecite{Shavrov1989,Shavrov1993,Tereshchenko2018}, however, the emphasis of these studies was on the spectrum of coupled magnetoelastic waves.
In those treatments, inhomogeneous deformations induced in the crystal by the magnetic background are accounted for, but the reverse effect of the elastic subsystem on magnetic ordering is ignored.
This approach is justified for the magnetization-induced strains, since the magnetoelastic coupling is much weaker than magnetic interactions.
In case of deformations resulting from an external stress, such an approach is generally not valid since essential transformations of the magnetic order may arise.

In this paper, the magnetic soliton lattice deformation by a uniaxial tensile stress applied perpendicular to the helicoidal axis is examined.
It turns out that the task is equivalent to a search of a magnetic configuration shaped simultaneously by an external magnetic field and a single-ion magnetic anisotropy that may be formulated in terms of the double sine-Gordon model (dSG).
The dSG model has been argued to be a model of several physical systems, such as the spin dynamics in the B phase of superfluid ${}^3$He \cite{Maki1976}, propagation of resonant ultrashort optical pulses through degenerate media \cite{Dodd1975}, nonlinear excitations in a compressible chain of dipoles \cite{Remoissenet1981}, and for soliton-like  misfit dislocations on the Au(111) reconstructed surface \cite{Stancioff1987}.
Emphasis of these studies was placed on solitary waves, or solitons, of infinite period, that are propagative solutions of a certain class of nonlinear partial differential equations \cite{Condat1983,Leung1983,Magyari1984,Pouget1984}.
Beyond this, the dSG model has found application in characterizing incommensurate (IC) structures in ferroelectrics \cite{Golovko1982,Iwabuchi1983,Techranchi1997} and ferromagnets \cite{Izyumov1983} that admit the Lifshitz gauge invariant.
The dSG model has also been employed to identify incommensurate phases of ferroelectrics in a non-zero electric field from the temperature dependent dielectric susceptibility \cite{Hudak1983}.

In magnetic systems, the different phases may be verified by using neutron scattering as a change of satellites spots in the diffraction data \cite{Izyumov1983}.
This paper introduces a way to identify the incommensurate phases arising in the double sine-Gordon model by means of Lorentz transmission electron microscopy.
In thin films of CrNb$_3$S$_6$, this experimental technique has proven to be effective in studies of magnetic order \cite{Togawa2012}, important features of magnetic chirality domains \cite{Togawa2015}, temperature dependence of the helical pitch \cite{Togawa2019}, and in the formation and movement of dislocations \cite{Paterson2019}.
With the example of deformations of the helicoidal order by mechanical stress, we demonstrate that Lorentz TEM imaging of the magnetic order through the real-space Fresnel technique~\cite{Chapman1999} allows identification of the magnetic phase.
We calculate a magnetic phase shift using the Fourier method, which has proved successful in other periodic magnetic structures \cite{Beleggia2000,Beleggia2001,Beleggia2003}, across a number of incommensurate structures in multiple magnetic phases originating from the combined effect of tensile strains and an external magnetic field.
We compare the calculations with experimental line profiles from Fresnel imaging of a strained sample of CrNb$_3$S$_6$ as a function of applied field and find good agreement.
These results establish strain-induced effects in CrNb$_3$S$_6$ and the presence of associated incommensurate phases in the material, with potential applications in fundamental research, magnonics, spintronics, and the emerging area of strain manipulated spintronics. 

This paper is organized as follows.
In Sec. II, we formulate a model of the chiral helimagnet subjected to tensile elastic strains.
Calculations of the Aharonov-Bohm induced electron phase shift for the different magnetic phases of the model are presented in Sec. III.
In Sec. IV, we discuss how to identify these magnetic phases by using the Fresnel technique of the Lorentz electron microscopy.
Results of experimental examination of strain effects in CrNb$_3$S$_6$ using TEM is reported in Sec. V wherein these findings are compared with theoretical predictions.
Our findings are summarized in Sec. VI.

\section{Theoretical Treatment}
The following is modelling of deformation of the magnetic chiral helix by an external magnetic field and tensile strains. 
The analysis focuses on parametrization of incommensurate phases and construction of a ``field-strain'' phase diagram.
The subject of the study is the chiral helimagnet of hexagonal symmetry, which is a prototype of the real compound CrNb$_3$S$_6$ \cite{Miyadai1983, Togawa2012, Ghimire2013, Togawa2015, Shinozaki2016, Wang2017, Togawa2019, Paterson2019}.

\begin{figure}[t]
    \centering
        \includegraphics[width=85mm]{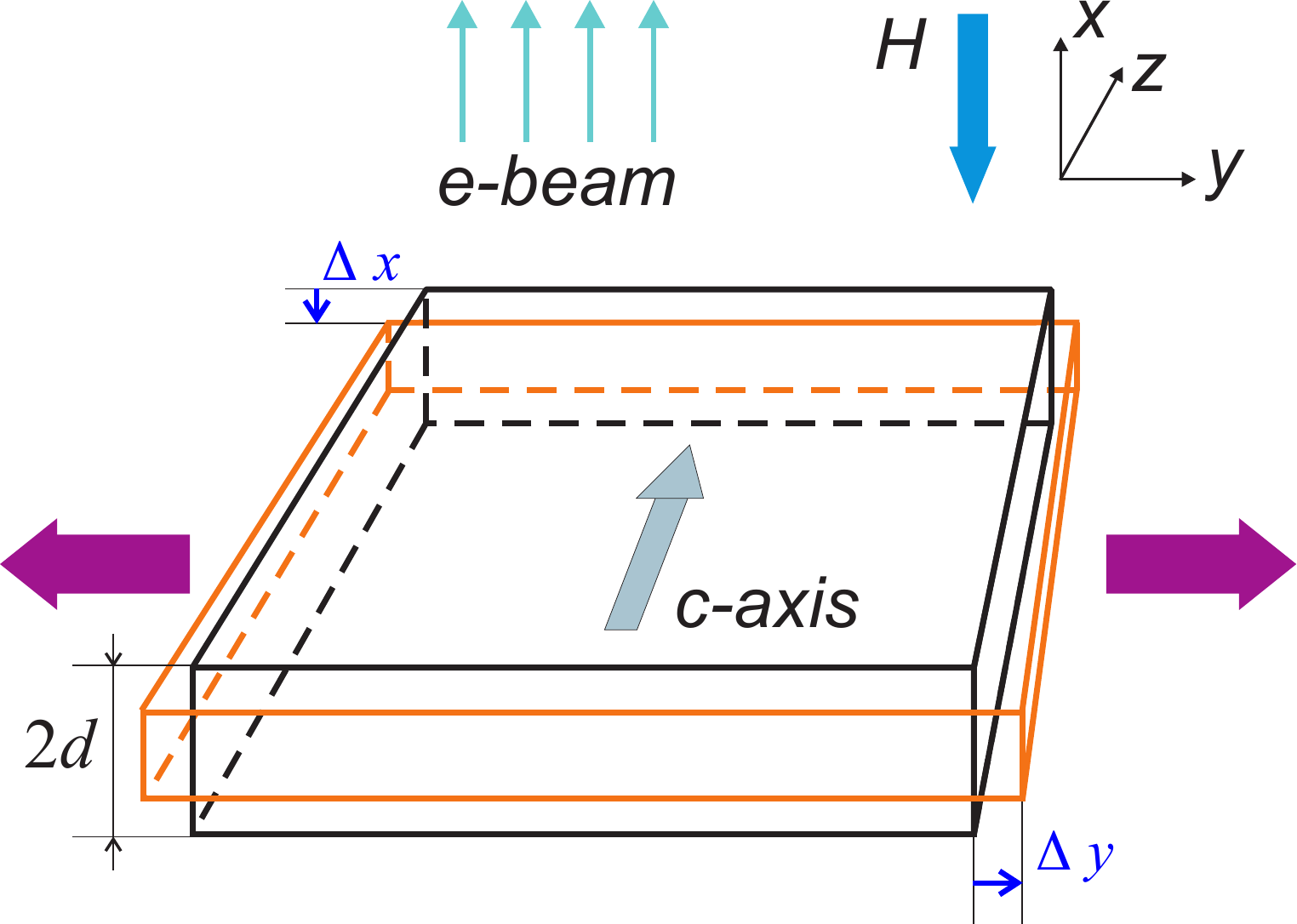}
        \caption{Schematic configuration of Lorentz microscopy of a sample of the chiral helimagnet of thickness $2d$ subjected to an external field $H$. A helix propagation vector is directed along the crystallographic $c$-axis. The tensile stress (large, magenta arrows) results in deformations of the nominal dimensions (black outline) by $\Delta x$ and $\Delta y$ along the $x$ and $y$ axes, respectively (orange outline).}
        \label{fig:fig1}
\end{figure}

The helicoidal magnetic order is characterized by the magnetiztion vector $\textbf{m}(\textbf{r},t)$ that determines  the total energy density, $\mathcal{F} =  \mathcal{F}_{\textrm{m}} + \mathcal{F}_{\textrm{me}}$.
The former component includes magnetic interactions,
$$
\mathcal{F}_{\textrm{m}} = \frac{J_{\perp}}{2}  \sum_{\alpha=x,y} \left(  \partial _{x_{\alpha}} \textbf{m}   \right)^2
+  \frac{J_{||}}{2}  \left( \partial_z \textbf{m} \right)^2
$$
\begin{equation}  \label{FMag}
- D \left( m_x \partial_z m_y - m_y  \partial_z m_x  \right)  -  H_x m_x,
\end{equation}
where the first two terms correspond to the exchange couplings in the plane perpendicular to the chiral axis, $J_{\perp}$, and along this axis, $J_{||}$. 
The Lifshitz invariant has the strength $D$ related with the Dzyaloshinskii-Moryia interaction along the chiral $c$-axis which is taken to lie along the $z$-axis, as shown in Fig.~\ref{fig:fig1}.
The last term describes the Zeeman coupling with the external magnetic field $\textbf{H}$, applied along the $x$-axis.
Hereinafter, we assume that $H_x \leq 0$.

The magnetoelastic energy density complies with symmetry of the 622 ($D_6$) point group of the hexagonal crystal \cite{Mason1954} 
$$
\mathcal{F}_{\textrm{me}} = \left( b_{11} - b_{12}  \right) \left( u_{xx} m^2_x + u_{yy} m^2_{y} + 2 u_{xy} m_x m_y  \right) 
$$
$$
+ \left( b_{13} - b_{12}  \right) \left( u_{xx} + u_{yy} \right) m^2_z + 
\left( b_{33} - b_{31}  \right) u_{zz} m^2_z 
$$
\begin{equation} \label{MEen}
+ 2 b_{44} \left( u_{yz} m_y m_z + u_{xz} m_x m_z \right),
\end{equation}
where $u_{ij}$ is the deformation tensor and $b_{ij}$ are the corresponding magnetoelastic constants. 

Taking note of the hierarchy $J_{\perp} >> J_{||}$ that exists in the real compound CrNb$_3$S$_6$ \cite{Shinozaki2016}, further simplifications may be reached if we neglect gradients of spin fluctuations in the plane perpendicular to the chiral axis.
In this case, the magnetization is confined within the $xy$-plane and modulated only along the $z$-direction, $\textbf{m} = M_s \left\{ \cos \varphi(z), \sin \varphi (z), 0 \right\}$, where $\varphi$ is the angle of the spin.
This approximation results in the following form of the magnetic energy
\begin{equation}
  \mathcal{F}_{\textrm{m}} = \frac12 J_{||}M^2_s  \left( \frac{d \varphi}{dz} \right)^2 - DM^2_s \left( \frac{d \varphi}{dz} \right) -  H_x M_s \cos \varphi. 
\end{equation}

We are going to examine deformation of the chiral soliton lattice subjected to stretch deformation arising from tensile strain along the $y$-axis (see Fig.~\ref{fig:fig1}), when only the components $u_{xx}$ and $u_{yy}$ are relevant.
This reduces the magnetoelastic interactions (\ref{MEen}) to the form
$$
\mathcal{F}_{\textrm{me}} = \left( b_{11} - b_{12}  \right) \left( u_{xx} m^2_x + u_{yy} m^2_{y}  \right)
$$ 
$$
= \mathcal{F}_0
+ \frac12 M^2_s  \left( b_{11} - b_{12}  \right) \left( u_{xx} -  u_{yy}  \right) \cos 2 \varphi 
$$
with $\mathcal{F}_0  = M^2_s  \left( b_{11} - b_{12}  \right) \left( u_{xx} + u_{yy}  \right)/2$. 
We see that the magnetoelastic interaction gives rise to the term $\cos 2 \varphi$, which has the same form as the second order single ion anisotropy term.
We note that the tensile deformations automatically bring about a non-zero $u_{zz}$ component, owing to the relationship $\Delta V/V = u_{xx}+u_{yy}+u_{zz}$, where $\Delta V/V$ is the relative change of volume.
The normal strain $u_{zz}$ is coupled with the magnetization $m_z$, as is evident from Eq.(\ref{MEen}).  
However, given the highly anisotropic magnetic nature of CrNb$_3$S$_6$, we assume that the conical spin order does not occur, and we thus ignore $u_{zz}$ in our model.

Minimization of the total energy by varying the angle $\varphi$ leads to the double sine-Gordon (dSG) model
\begin{equation} \label{DSG}
\frac{d^2 \varphi}{dz^2} + \frac12 b_1 \sin \varphi + b_2 \sin 2 \varphi =0,
\end{equation}
where $b_1 = - 2H_x/(J_{||}M_s)$ and $b_2= \left( b_{11} - b_{12}  \right)(u_{xx}-u_{yy})/J_{||}$.
Here, $b_1 > 0$ as long as $H_x \leq 0$.  
  
The dSG equation is well studied \cite{Iwabuchi1983,Izyumov1983}, and has two uniform solutions with $\cos \varphi =-1$ and $\cos \varphi = -b_1 /(4 b_2)$ that correspond to the commensurate (C) phases 1 and 2, respectively.
The 1C-phase exists for $b_1 > 4b_2$ and the 2C-phase exists for $b_1<4b_2$ provided that $b_1>0$.  
In addition to the commensurate solutions, there are two spatially nonuniform ones that describe incommensurate phases with one type (1s) and two types (2s) of the solitons \cite{Izyumov1983}.

\begin{figure}[t]
    \centering
        \includegraphics[scale=0.95]{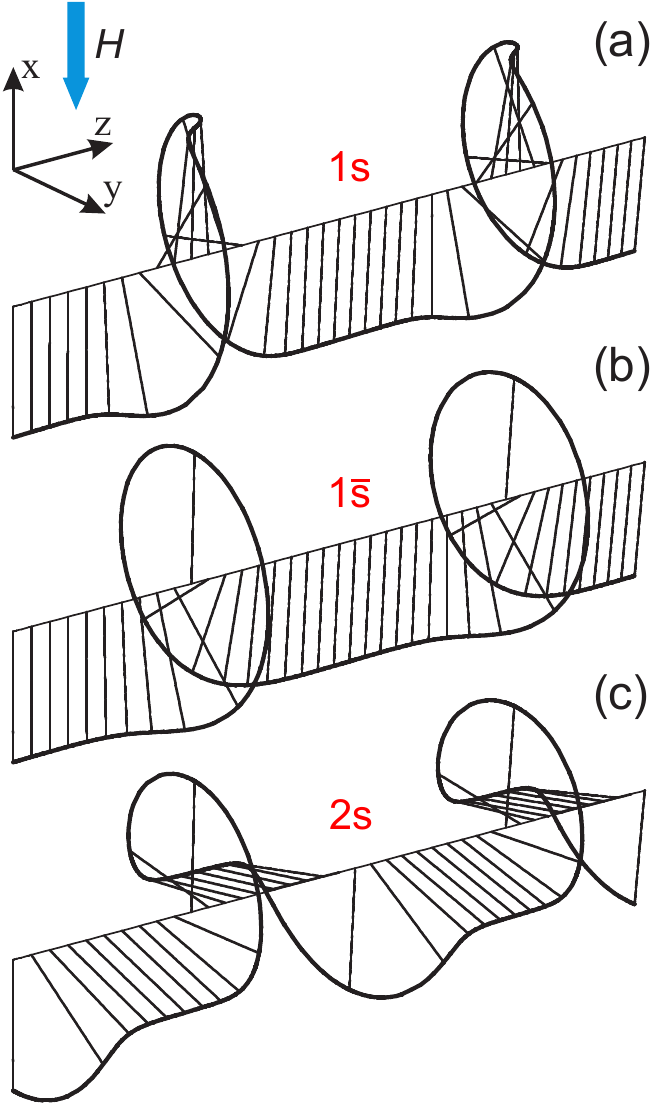}
        \caption{3-D profiles of magnetization for the incommensurate phases of the double sine-Gordon model: (a, b) the 1s-phase, and (c) the 2s-phase. The 1s and 1$\bar{{\textrm{s}}}$ plots correspond to different parameterisations of the same phase (see main text for details) and are differentiated here by the emergence of ferromagnetic areas aligned along the positive $x$-axis \emph{inside} the kinks of the 1s plot.}
        \label{fig:fig2}
\end{figure}

The 1s-solution is given by
\begin{equation}  \label{RegI}
  \cos \varphi = \frac{2x_1 \textrm{sn}^2 \left( \bar{z} \right) + 1 -x_1}{2 \textrm{sn}^2 \left( \bar{z} \right) - 1 + x_1},
\end{equation} 
where $\textrm{sn}(\ldots)$  is the Jacobi elliptic function of the dimensionless coordinate $\bar{z} = 2K(z-z_0)/L_0$, with $K$ being the elliptic integral of the first kind.
The point $z_0$ results from the condition $\varphi (z_0) = \pi$ [see Eq.(\ref{GenInt}) in Appendix A].
The elliptic modulus
\begin{equation}  \label{k2phase1}
  \kappa^2 = \frac{2(x_2-x_1)}{(x_2+1)(1-x_1)}
\end{equation} 
and the period of the solution 
$$
L_0 = \frac{4K}{\sqrt{2 b_2 (x_2+1)(x_1-1)}}
$$
are determined by the parameters $x_1<-1$ and $x_2>1$  that  depend on the constants $b_{1,2}$ and an integration constant of the dSG  equation [see Eq.(\ref{x12}) in Appendix A].
The same solution is valid for the relationship $x_2<x_1<-1$.
To distinguish between these cases we denote the latter by 1$\bar{{\textrm{s}}}$.
In this phase, the magnetic moments rotate around the $z$-axis but are mostly directed along the direction of the external magnetic field [see Figs.~\ref{fig:fig2}(a) and \ref{fig:fig2}(b)].
Due to spin alignment in the almost ferromagnetically ordered areas, this spin arrangement is related to the commensurate 1-phase. 

The corresponding energy per unit length [Eq.(\ref{EnS1})] is expressed in terms of the complete integrals of the first, second ($E$) and third kind,
$$
 \Pi \left( \alpha^2, \kappa \right) = \int^{\pi/2}_0  \frac{d\varphi}{\left( 1 - \alpha^2 \sin^2 \varphi \right) \sqrt{1-\kappa^2 \sin^2 \varphi}},  
$$
where $\alpha^2 = 2/(1-x_1)$.
The parameter $x_1$ corresponds to the integration constant of Eq.(\ref{DSG}) and must be obtained from a minimum of the system energy.
This leads to the equation to find $x_1$:
$$
 - \frac12 \pi q_0 \sqrt{2b_2 \left( x_2+1 \right) \left( x_1 -1 \right)}
+ b_2 K \left(1-x^2_1 \right) 
$$
\begin{equation}\label{Optx1}
+ b_2 E \left( x_2 +1 \right) \left( x_1 -1 \right)
+ b_2 \Pi  \left( x_1 + x_2 \right) \left( x_1 + 1 \right) = 0,
\end{equation} 
where $x_2 = -x_1 -b_1/2b_2$, and $q_0=D/J_{||}$ is the wavenumber of the helimagnetic structure under zero field and zero stress.

The 2s-solution is defined as (see Appendix A)
\begin{equation}  \label{RegIII}
  \cos \varphi = \frac{l-\textrm{cn} \left( \bar{z} \right)}{1- l \textrm{cn} \left( \bar{z} \right)},
\end{equation} 
where $\textrm{cn}(\ldots)$ is the Jacobi elliptic function  of the dimensionless coordinate $\bar{z} = 4K(z-z_0)/L_0$. It has the period  
$$
L_0 = 4 K \sqrt{\frac{2l}{b_1 \left( l^2 - 1 \right)}}
$$
and depends on the elliptic modulus 
\begin{equation}  \label{kappa2s}
k^2  = \frac{l(l+4b_2/b_1)}{l^2-1},
\end{equation} 
which are both specified by the parameter $l$.
This parameter is linked to $b_1$ and $b_2$ by the relationship $c/b_1 = -b_2/b_1 - l/2 -1/(2l)$, where $c$ is a constant of integration of Eq.(\ref{DSG}). 
The parameter $l$ may vary in the range $-b_1/(4b_2) <l<0$ for $b_2/b_1>1/4$ and in the range $-4b_2/b_1<l<0$ for  $b_2/b_1<1/4$.
Minimization of the energy per unit length [see Eq.(\ref{EnS2})] fixes $l$ through the relationship 
$$
- \pi q_0 \sqrt{\frac{b_1}{2}\left( l - 1/l \right)} + b_1 \left( \Pi - K \right) \left( l + {b_1}/{4b_2} \right) 
$$
\begin{equation} \label{Optl}
+ b_1 \left( l - {1}/{l} \right) E =0, 
\end{equation}
where $\Pi$ includes the elliptic characteristic $\alpha^2 = -4lv$ with $v=b_2/b_1$.

In the 2s-phase, there are two energetically equivalent directions of the magnetic moments [Fig.~\ref{fig:fig2}(c)].
Although the moments rotate around the $z$-axis, they are mainly aligned along the directions with $\cos \varphi = -b_1 /(4 b_2)$ (corresponding to the angle of the commensurate 2-phase, discussed above), effectively spreading the 2$\pi$ rotations across some distance.
This occurs in order to balance competition between the external magnetic field and an effective second order anisotropy arising from the magnetoelastic coupling.

\begin{figure}[t]
    \centering
        \includegraphics[width=85mm]{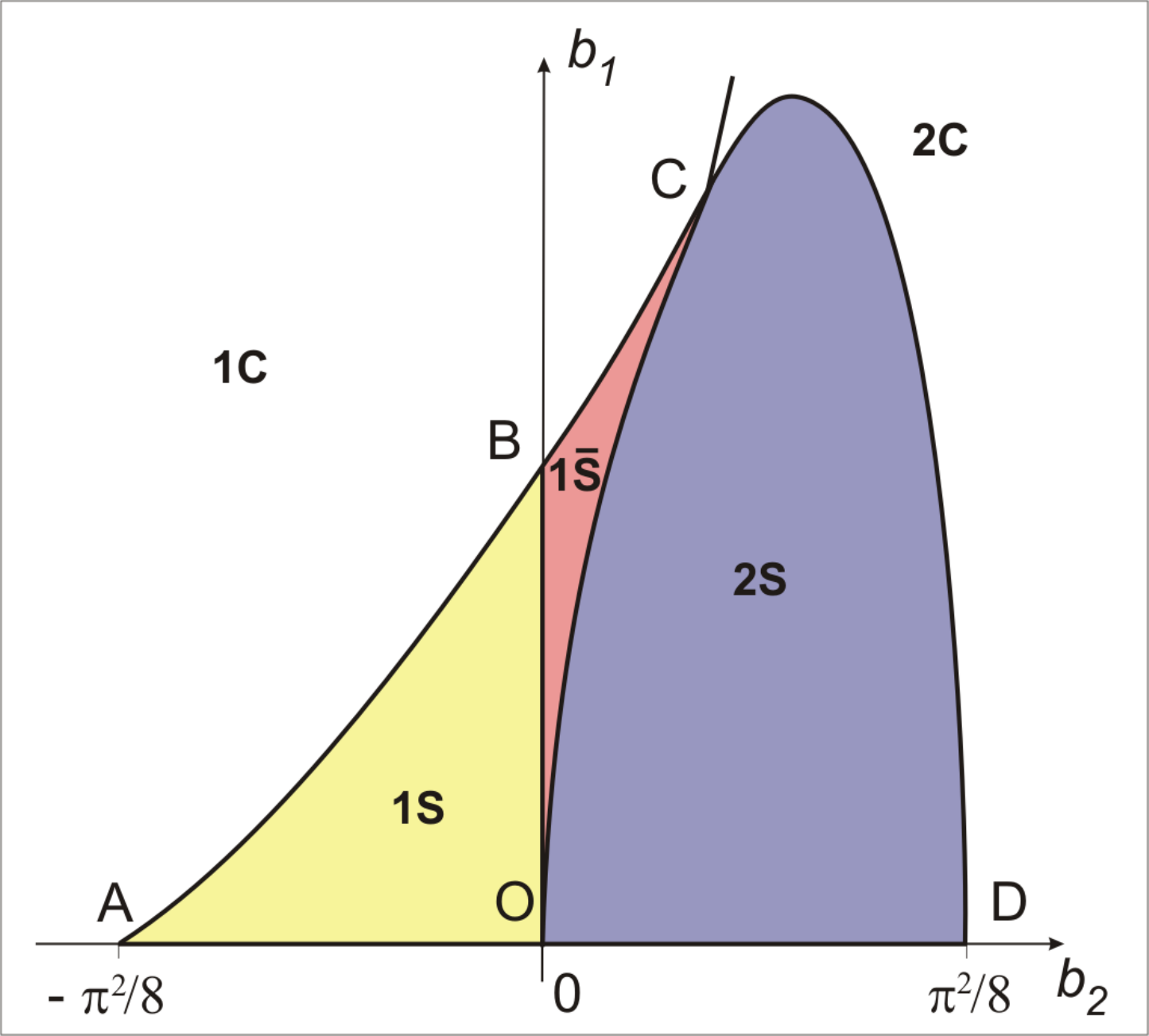}
        \caption{The phase diagram in the ($b_2$-$b_1$) plane. The scale along the axes $b_{1,2}$ is chosen in units of $q^2_0$. The parameters $b_2$ and $b_1$ correspond to the elastic deformation and applied field, respectively. Examples of characteristic spin textures for the different phases are depicted in Fig.~\ref{fig:fig2}. The commensurate 1-phase (1C) and 2-phase (2C) are separated by the line $b_1/b_2=4$.}
        \label{fig:fig3}
\end{figure}

The phase diagram corresponding to the thermodynamical potential $\mathcal{F}$ on the plane formed by the parameters $b_2$ and $b_1$, which are associated with the elastic deformation and applied field, respectively, is shown in Fig.~\ref{fig:fig3}.\footnote{In general, the sign of the magnetoelastic constants in unknown. However, we demonstrate later that the 1s phase appears to correspond to the case of tensile strain in CrNb$_3$S$_6$.}
It is identical to that given in Ref.~\onlinecite{Izyumov1983}, which was built from a free-energy Ginzburg-Landau functional including a second order magnetic anisotropy and an external magnetic field.
The lines presented in the phase diagram are determined by the relations:
\begin{flushleft}
(AB)-line:
\end{flushleft}
\begin{equation} \label{AB}
  - \pi q_0 \sqrt{2b_1} + 2b_1 \sqrt{1-4v} + \frac{b_1}{\sqrt{|v|}} \textrm{arcsinh}(2\sqrt{|v|}) = 0,
\end{equation}
\begin{flushleft}
(BC)-line:
\end{flushleft}
\begin{equation} \label{BC}
  - \pi q_0 \sqrt{2b_1} + 2b_1 \sqrt{1-4v} + \frac{b_1}{\sqrt{v}} \textrm{arcsin}(2\sqrt{v}) = 0,
\end{equation}
\begin{flushleft}
(OC)-line:
\end{flushleft}
\begin{equation} \label{OC}
 b_1 = 2 q_0  \sqrt{2b_2},
\end{equation}
\begin{flushleft}
(CD)-line:
\end{flushleft}
\begin{equation} \label{CD}
 \frac{\pi q_0}{b_1} \sqrt{2b_2} = \sqrt{16v^2-1} +  \textrm{arcsin} \left( \frac{1}{4v} \right).
\end{equation}
It should be noted that the OB line divides the area of the 1s-phase where different forms of $x_{1,2}$ parametrization are used, and does not correspond to any phase transition.
The boundaries between the incommensurate and commensurate phases, i.e., the AC and CD lines, are related with  continuous second-order phase transitions of the nucleation type, according to de Gennes classification, \cite{Gennes1975} and may be derived from the limit $\kappa^2 \to 1$.
In this limit, the definitions (\ref{k2phase1}) and (\ref{kappa2s}) impose additional constraints on the parameters $x_1$, $x_2$, or $l$ as functions of $b_1$, $b_2$.
Given these connections, similarly treating Eqs.(\ref{Optx1}) and (\ref{Optl}) at $\kappa^2 \to 1$ leads to Eqs.(\ref{AB}), (\ref{BC}), and (\ref{CD}), respectively.
The transition across the OC-border relates to a continuous transformation from one incommensurate phase to another that happens at $\kappa^2=0$.
In Ref. \cite{Golovko1982} it has been argued that this boundary is not a phase transition line, since any derivative of thermodynamic potential is continuous at the boundary \cite{OurComment}.
Repeating the above procedure, i.e., by addressing Eqs.(\ref{Optx1}) and (\ref{Optl}) subject to new links of the limit, one may deduce Eq.(\ref{OC}).

The coordinates of the points A, B, C and D in the $(b_2,b_1)$-plane are $q^2_0 \left( -\pi^2/8, 0\right)$, $q^2_0 \left( 0, \pi^2/8\right)$,  $q^2_0 \left( 0.5,  2 \right)$ and $q^2_0 \left( \pi^2/8, 0\right)$, respectively.
We highlight these specific points because the exchange coupling $J$ and the magnetoelastic constant $b_{11}-b_{12}$  may be found from comparison with critical values of the incommensurate-commensurate phase transition either at zero strain or at zero magnetic field.

\section{Simulated Lorentz Profiles}
We examine in detail below the magnetic phase shift imparted to an electron wave passing through a specimen with an incommensurate magnetic order of the dSG model.
For the rectangular geometry sketched in Fig.~\ref{fig:fig1}, the magnetic phase shift, based on the Aharonov-Bohm effect, is given by \cite{Fukuhara1983}
\begin{equation} \label{Mshift}
\phi(z) = - \frac{e}{\hbar} \int_{l} A_x (\textbf{r}) d\textbf{r},
\end{equation}
where the line integral is performed within the specimen for samples with no stray field, $\textbf{A}$ is the magnetic vector potential inside the sample, $e$ is the absolute value of the electron charge and $\hbar$ is the reduced Plank constant.
The specimen has the form of a thin slab of constant thickness $2d$.
The electron beam passes through the film, parallel to the $x$-axis, and the object in the $yz$-plane lies perpendicular to the incident beam, as shown in Fig.~\ref{fig:fig1}.
It is supposed that the magnetic phase shift depends only on the $z$ coordinate and does not change in the $xy$-plane lying perpendicular to the helicoidal axis.
Since the recorded phase is simply related with the in-plane magnetisation in the sample
$$
M_y = -\frac{\hbar}{2e \mu_0 d} \partial_z \phi,  \qquad 
M_z =  \frac{\hbar}{2e \mu_0 d} \partial_y \phi,
$$
it is easy to see that  the $y$-component of the sample magnetization $M_y(\textbf{r}) = M_0 \sin \varphi (\textbf{r})$ alone determines   the $x$-component of the magnetic vector potential
\begin{equation}
A_x(\textbf{r}) = \frac{\mu_0}{4\pi} \int M_y (\textbf{r}') \frac{(z-z')}{|\textbf{r}-\textbf{r}'|^3} d \textbf{r}'.
\end{equation}

It is convenient to use the Fourier transform of this expression
$$
A_x(\textbf{k}) = - i \mu_0 \frac{k_z}{\textbf{k}^2} M_y (\textbf{k}),
$$ 
where $ M_y (\textbf{k}) = M_0 \int d \textbf{r} \sin \varphi (\textbf{r}) e^{-i \textbf{k}\textbf{r}}$.   

Direct calculation for the specified geometry results in 
\begin{equation} \label{Myk}
M_y(\textbf{k}) = 4\pi M_0 \delta(k_y) \frac{\sin(k_xd)}{k_x} \int^{\infty}_{-\infty} dz e^{-ik_z z}\sin \varphi(z),
\end{equation}
where, for simplicity, we assume that the slab plane is infinite.  

Then, the magnetic phase shift is given by 
$$
\phi(z) = - \frac{e}{\hbar} A_x(k_x=0,z)
$$
$$
=  - \frac{e}{\hbar} \int^{\infty}_{-\infty} \frac{dk_y}{2\pi}  \int^{\infty}_{-\infty} \frac{dk_z}{2\pi}  A_x(0,k_y,k_z) e^{ ik_z z}.
$$

Next, we give results for the magnetic phase shift separately for each of the phases.
In the following section, we use these equations to examine the Fresnel contrast that the different magnetic phases produce.

\emph{\textbf{1s-phase.}}
In the case of the 1s-phase, with the parameters $x_1<-1$ and $x_2>1$, evaluation of the magnetization (\ref{Myk}) can be carried out with the aid of the Fourier series for $\sin \varphi (z)$ (see Appendix C).
Then, the ultimate result for the magnetic phase shift is as follows
\begin{equation}  \label{PhaseShift1}
\phi(z) = - \frac{2e\mu_0 M_0 d}{\hbar \sqrt{-2b_2}}  \left[
  \textrm{am} (\bar{z}+b) - \textrm{am} (\bar{z}-b) - \dfrac{\pi}{K}b
\right],
\end{equation}
where $\textrm{am} (\ldots)$ is the Jacobi's amplitude function and $b$ is specified by Eq.(\ref{bRI}).

In the case of $x_2<x_1<-1$, the result
$$
\phi(z) = \dfrac{2e\mu_0 M_0 d}{\hbar \sqrt{2b_2}}  \left\{  i \textrm{am} \left[ \bar{z} - K + i (K'-b) \right]
\right.
$$
\begin{equation}  \label{Phi2}
\left.
-  i \textrm{am} \left[ \bar{z} - K -  i (K'-b) \right] + \frac{\pi}{K} \left( K'-b\right)
\right\}
\end{equation}
may be obtained by application of the same technique. 
Here, $b$ is determined by Eq.(\ref{bRII}) and $K'$ denotes the complete elliptic integral of the first kind with the complementary elliptic modulus $\kappa'= \sqrt{1-\kappa^2}$.

\emph{\textbf{2s-phase.}}
For the 2s-phase, derivation of the Fourier series for the transverse magnetization component is covered in Appendix D.
Using the Fourier transformation in the general scheme of magnetic phase shift calculations, we obtain the overall result
$$
\phi(z)  = 2 \frac{e\mu_0 M_0 d}{\hbar \sqrt{2b_2}} 
\left\{
 -\frac14 \ln \left( \dfrac{1-\kappa \textrm{cd} \left[ \bar{z} + i (K'-b)\right]}{1+\kappa \textrm{cd} \left[ \bar{z} + i (K'-b)\right]}  \right)
\right.
$$
$$
-\frac14 \ln \left( \dfrac{1-\kappa \textrm{cd} \left[ \bar{z} - i (K'-b)\right]}{1+\kappa \textrm{cd} \left[ \bar{z} - i (K'-b)\right]}  \right)
+ \frac{\pi}{2K} \left( K' - b \right)
$$
\begin{equation} \label{PhaseShift2s}
\left.
- \frac{i}{2} \textrm{am} \left[ \bar{z} - i(K'-b) \right]
+ \frac{i}{2} \textrm{am} \left[ \bar{z} + i(K'-b) \right]
\right\}.
\end{equation}
A link between the parameters $b$ and $l$  is conditioned by Eq.(\ref{bParam}) in Appendix~D.

\begin{figure*}[t]
    \centering
    \includegraphics[width=17cm]{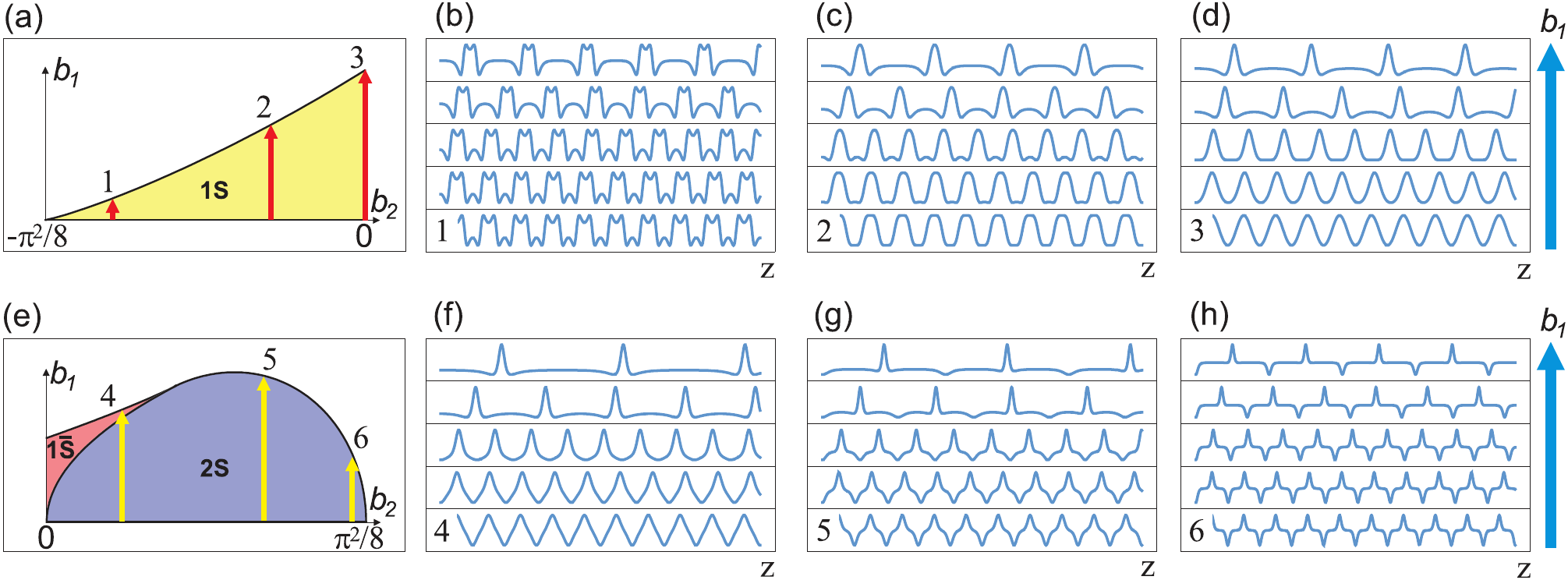}
    \caption{Upper panels: profile evolution of $\phi^{''}(z)$  with the magnetic field for the 1s magnetic phase along the selected directions of the phase diagram indicated by the red arrows in (a): (b) at large ($b_2=-0.8 \pi^2/8$), (c) at small ($b_2=-0.3 \pi^2/8$) and (d) at zero elastic deformations. 
    Lower panels: equivalent plots for the 1$\bar{{\textrm{s}}}$ and 2s phases shown in (e): (f) at small ($b_2=0.24 \pi^2/8$), (g) at intermediate ($b_2=0.67 \pi^2/8$), (h) and at large ($b_2=0.96 \pi^2/8$)  elastic deformations. The $b_2$ values are given in units of $q^2_0$. 
    The values of $b_1$ for all plots are chosen as (from bottom to top) 0.05, 0.30, 0.65, 0.99 and 0.9999 from a critical value of the IC-C phase transition at a given $b_2$ value.}%
    \label{fig:fig4}
\end{figure*}

\section{Identification of the dSG phases by TEM}
It is instructive to consider the contrast that would be obtained from Lorentz TEM imaging of the different phases outlined in the previous section.
Electron holography \cite{Tonomura_electron_holography} provides access to the phase shift imparted on the transmitted beam due to components of the integrated induction lying perpendicular to the beam.
The differential phase contrast (DPC) technique essentially maps the first derivative of the magnetic phase change, similar to the Foucault imaging technique \cite{Chapman1983, Chapman1990}.
In the Fresnel technique, contrast sensitive to the second derivative of the phase is obtained by defocusing the image.
It is routinely used to image domain walls \cite{McVitie2006}, and it is with this technique that remainder of this section is concerned.

While Fresnel imaging of domain structures is generally considered to be a non-linear imaging mode, at small defocus values the Fresnel image intensity, $I$, is linear in the second derivative of the phase \cite{McVitie2006} and may be used directly in a quantitative manner \cite{Beleggia2004}: 
\begin{equation}
 I (\textbf{r}) \approx  1- \frac{\Delta f \lambda}{2\pi} \nabla^2_{\perp}   \phi (\textbf{r}) =  1 - \frac{\Delta f \lambda}{2\pi} \phi ''(z)
\end{equation}
with $\lambda$ the electron wavelength, $\Delta f$ the defocus distance, and $\nabla^2_{\perp}$ the Laplacian relating to coordinates perpendicular to the beam.

Next, we examine the Fresnel contrast for different regions of the phase diagram with reference to Fig.~\ref{fig:fig4}.
For each panel row in the figure, the second to fourth columns depict the simulated profiles at positions along the arrows in the first column, with the field strength increasing from bottom to top.
At the outset, we examine a situation in the absence of elastic deformations.
In this case, we obtain the sine-Gordon model whose behavior has been well explored by TEM \cite{Togawa2012,Togawa2015,Togawa2019,Paterson2019}.
At small fields, the magnetic order is close to a helix and the profile of the second derivative of the phase imparted on the transmitted beam approximates a sinusoidal waveform [bottom row of Fig.~\ref{fig:fig4}(d)].
With an increase in the field [moving up the rows in Fig.~\ref{fig:fig4}(d)], the main peaks corresponding to 2$\pi$ rotations of the spins, narrow and additional low magnitude peaks develop.
In achieving the incommensurate-commensurate phase transition, these additional peaks broaden and transform into plateaus.
This situation reflects modification of the soliton lattice by the external magnetic field, namely a growth of commensurate, forced ferromagnetic regions (FFM) in order to lower the Zeeman energy.

Switching on small deformations gives rise to an effective anisotropy field parallel to the magnetic one, but unlike the latter, it prefers magnetization directions both along and against the magnetic field [see Fig.~\ref{fig:fig2}(a)].
As a result, this leads to a more rapid broadening of the main peaks and the shallow peaks associated with the FFM regions [Fig.~\ref{fig:fig4}(c)].

At stronger deformations, the induced anisotropy field dominates, changing drastically the $\phi^{''}(z)$ profile, causing bifurcation of the main peaks and the appearance of additional peaks between the main ones, resulting in a symmetrical `up-up-down-down' structure [Fig.~\ref{fig:fig4}(b)].
Increasing the applied field breaks the symmetry due to formation of FFM regions that are visible as broad plateaus as the IC-C phase transition between the 1s and 1c phases  is approached.

A specific feature of the 1$\bar{{\textrm{s}}}$-phase evolution is that it is preceded by the 2s-phase at small magnetic fields [arrow 4 in Fig.~\ref{fig:fig4}(e)].
With the growth of the magnetic field, the triangular waveform of the 2s phase transforms continuously to a profile with sharp peaks separated by plateaus, however, in contrast to the 1s phase, without additional peaks [Fig.~\ref{fig:fig4}(f)].
Another characteristic feature is that these peaks are narrower than those of the 1s phase near the IC-C phase transition of the 1$\bar{{\textrm{s}}}$-1c type.

A change of the electron phase second derivative under the effect of a magnetic field occurs entirely inside the 2s-phase at stronger deformations [arrows 5 and 6 in Fig.~\ref{fig:fig4}(e)].
At small fields, the strain induced anisotropy field prevails, so there is a non-sinusoidal symmetrical profile [Figs.~\ref{fig:fig4}(g) and \ref{fig:fig4}(h)].
However, only one kind of peaks survives as the field increases, namely, that which corresponds to fast rotating kinks with advantageous Zeeman arrangement of moments.
The broad plateau areas have characteristic dips whenever magnetization reaches full saturation.
These plateaus are a manifestation of the commensurate magnetic order of the 2c-type, i.e. the canted phase.

\section{Experimental Results}

\begin{figure}[t]
    \centering
        \includegraphics[width=8.5cm]{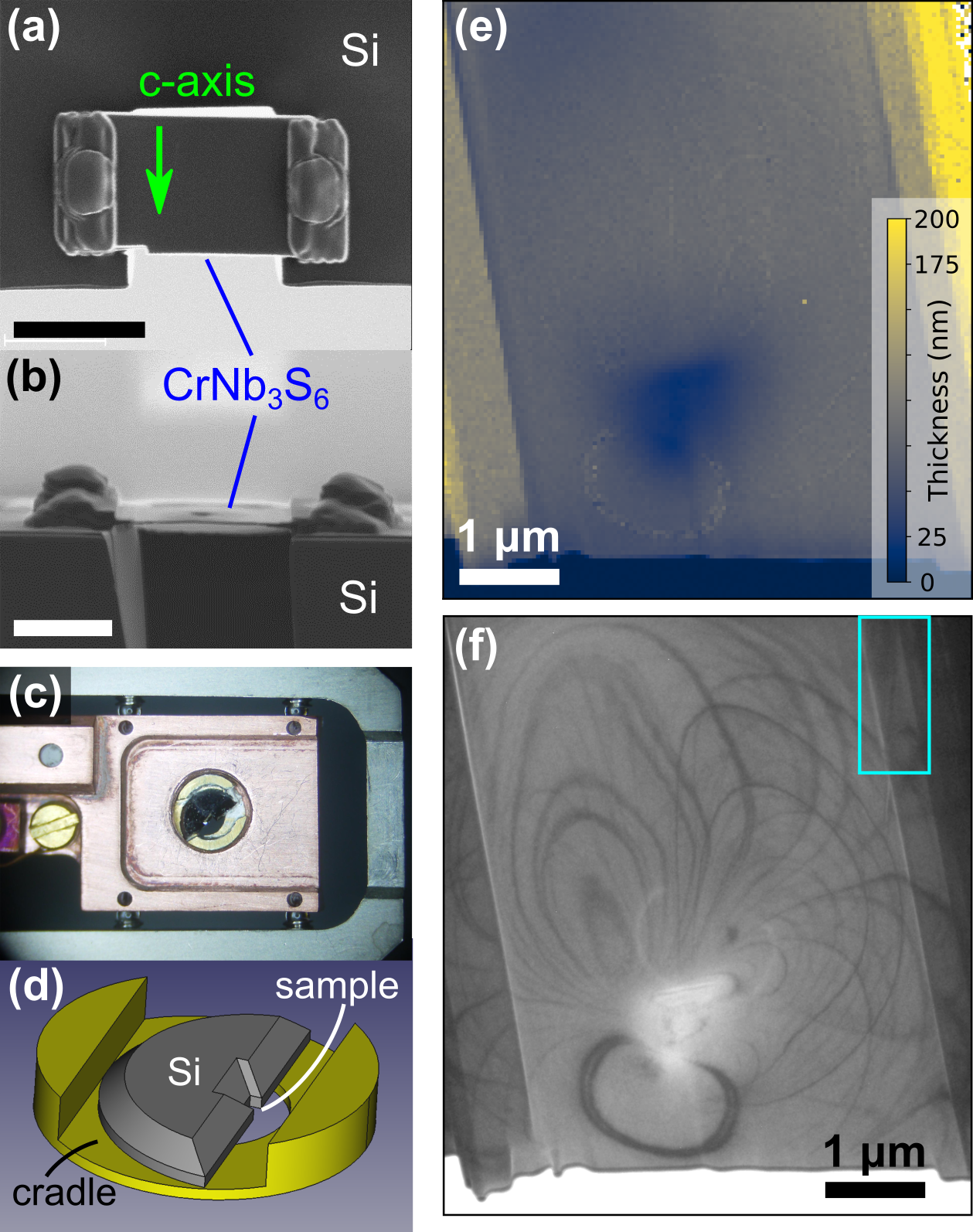}
        \caption{CrNb$_3$S$_6$ sample preparation and characterisation.
        (a, b) Top and side view SEM images of the sample mounted on a Si support, pre and post thinning, respectively.
        The scale bars in (a) and (b) are 4.5~$\muup$m.
        (c) Sample mounted in a Gatan HC3500 sample holder in a custom brass sample `cradle'.
        (d) Schematic of the sample cradle used to isolate the sample from strain during mounting.
        (e) Thickness map from electron energy loss spectroscopy performed in scanning mode TEM (STEM), analysed following Iakoubovskii method \cite{Iakoubovskii2008} implemented in the fpd library \cite{fpd} using a density of 5.03 g/cm$^3$ \cite{Tereshchenko2018}, giving an inelastic mean free path of 126~nm.
        The convergence and collection semi-angles were 29~mrad and 36~mrad, respectively. 
        (f) CTEM image of the same area displayed in (e) showing bend contours and (cyan rectangle) the region under study.
        }
    \label{fig:fig5}
\end{figure}

In order to experimentally observe strain effects discussed in the theory section of this work, we examine the prototypical chiral helimagnet CrNb$_3$S$_6$ by TEM.
We adopt a similar approach to that of Shibata {\it{et al}}. \cite{Shibata2015} by aiming to induce tensile strain upon cooling through differential coefficients of expansion of the sample mounted on a Si support.
An electron transparent focused ion beam prepared cross-section of the crystal was mounted over a slot in a Si support structure [Figs.~\ref{fig:fig5}(a) and \ref{fig:fig5}(b)] and the sample cooled to 106~K, below the critical temperature of $\sim$127~K.
A custom intermediate sample holder [Fig.~\ref{fig:fig5}(d)] was used to avoid additional strain arising from mounting in a Gatan HC3500 sample holder [Fig.~\ref{fig:fig5}(c)].
During cooling, the lower thermal expansion rate of the Si support nominally causes tensile strain to form in the sample in the direction across the Si gap, corresponding to a direction perpendicular to the chiral $c$-axis.
However, due to non-uniform thickness of the lamella [Fig.~\ref{fig:fig5}(e)] and sample warping during thinning [Fig.~\ref{fig:fig5}(b)], additional localised strain terms may arise and, indeed, the precise magnetic order varied across the sample.
Curvature of the sample is also visible in the conventional TEM (CTEM) image of Fig.~\ref{fig:fig5}(f) through the presence of bend contours, diffraction effects that create intensity variations due to gradual variation in the crystal alignment with respect to the beam.
Because of these effects, unfortunately, we cannot quantify the strain in the sample from the nominal structural and thermal properties of the sample (where they are known).
However, we can estimate the strain from comparison with theory, as discussed at the end of this section.

To characterise the field dependence of the magnetic configuration in the sample, it was imaged in Fresnel mode in an JOEL ARM~200cF TEM equipped with a cold field emission gun operated at 200~kV \cite{mcvitie2015_magtem}.
An external field was applied perpendicular to the sample $c$-axis by varying the current through the objective lens, from the remnant field (104~Oe) until the sample was fully field-polarised ($\sim$2000~Oe), while maintaining a defocus of $\sim$300 ~$\muup$m.

\begin{figure*}[t]
    \centering
        \includegraphics[width=16cm]{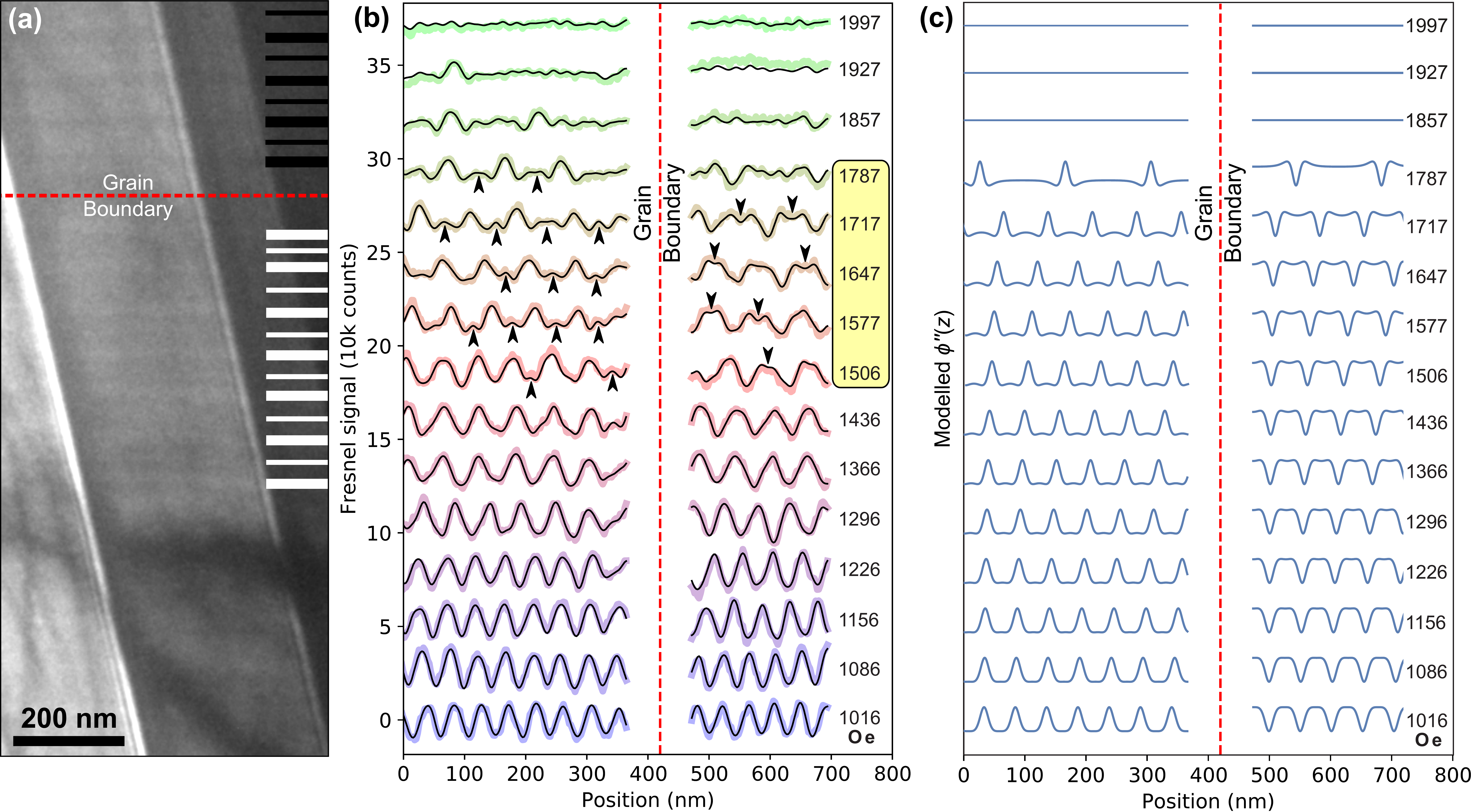}
        \caption{Comparison of experimental strain induced CrNb$_3$S$_6$ Fresnel contrast with simulated data.
        (a) Lorentz TEM image at an applied field of 1728~Oe from the region under study in Fig.~\ref{fig:fig5}(f).
        The overlaid black and white lines highlight the weak magnetic contrast.
        The thickness of the middle stripe is 126~nm.
        (b) Average experimental Fresnel line intensity profiles as a function of the applied field, showing additional peaks and troughs (prominent examples are arrowed) in the field range of $\sim$ [1500, 1800]~Oe.
        The black lines are [20, 200]~nm period bandpass filtered versions of the averaged data (coloured lines).
        The red dashed lines in (a) and (b) mark the approximate location of the crystallographic grain boundary.
        (c) Simulated contrast for the elastic deformation parametrised by $b_2= -0.01 \cdot \pi^2 q^2_0/8$ that corresponds to the 1s phase, showing very similar features to those shown in (b).
        This $b_2$ value is related with the upper limit $b^{\ast}_{1} = 0.976 \cdot \pi^2 q^2_0/8$ as follows from Eq.(11).
        For comparison with the experimental data, $b^{\ast}_{1}$ was set as 1787.1~Oe.
        }
    \label{fig:fig6}
\end{figure*}

Figure~\ref{fig:fig6}(a) shows a Fresnel image of the region under study, located at a crystallographic grain boundary defining regions of opposite chirality (marked by the red dashed line) at an intermediate field of 1728~Oe.
Solitons are visible as narrow dark or bright lines (marked by the black and white overlays), depending on the chirality of the grain \cite{Togawa2015}.
Across the imaged area, the solitons remain parallel to one another as a result of the c-axis being aligned \cite{Togawa2015}.
Importantly, between the solitons are narrower lines of the same intensity sign which are not observed in strain-free samples (see Fig.~3 of Ref.~\cite{Togawa2015} for examples).

Averaged experimental line profiles from imaging the sample in Fresnel mode as a function of applied field are shown by the coloured lines in Fig.~\ref{fig:fig6}(b).
The overlaid black lines are the same data Fourier filtered to mainly remove high frequency components primarily deriving from noise.
The same features of the profiles are present on both sides of the grain boundary, except for an inversion of the contrast.
At low applied field strengths, the soliton profile is approximately sinusoidal but, as the field strength increases, additional peaks (or troughs) become apparent (arrowed), starting at values of $\sim$1500~Oe and extending to $\sim$1800~Oe, before the sample becomes field polarised at $\sim$2000~Oe.
Figure~\ref{fig:fig6}(c) shows the field dependence of the simulated contrast of the 1s phase, with parameters chosen to match the experimental profiles.
While the contrast transfer function of the main imaging lens in the experiment reduces the high spatial frequency components in the experimental profiles, the general features and evolution of the profiles with field match well those of the calculated profiles, including the appearance of additional peaks while the lattice is still relatively dense.

Finally, from our identification of the approximate point in the phase diagram for our experiment, $b_2/b_1 \approx -10^{-2}$, we can estimate the strain from known CrNb$_3$S$_6$ values of the magnetization $M_s=131.5 \, \textrm{kA} \cdot \textrm{m}^{-1}$ and the magnetoelastic constants $\left( b_{11} - b_{12} \right) M^2_s \sim 1 \,  \textrm{MPa}$~\cite{Tereshchenko2018} through
\begin{equation}
    u_{xx} - u_{yy} =  - \frac{2 \mu_0 H_x M_s}{\left( b_{11} - b_{12} \right) M^2_s} \cdot \frac{b_2}{b_1},
\end{equation}
where $\mu_0 = 4\pi \cdot 10^{-7} \, \textrm{H}/\textrm{m}$ is the vacuum permeability.
Given the choice $H_x = -1700 \cdot \left( 4\pi \right)^{-1} \cdot 10^3  \, \textrm{A} \cdot \textrm{m}^{-1}$,  that corresponds to $-1700 \, \textrm{Oe}$, we obtain an estimate of the strain in our sample of $u_{yy} - u_{xx} \approx 4.47 \cdot 10^{-4}$, which is of the same order of magnitude as observed in similar experiments.~\cite{Shibata2015}

\section{Conclusions}
In summary, we have studied the influence of an external tensile force on a monoaxial helimagnets having a helical magnetic order due to competition between magnetic anisotropies in chiral crystals with Dzyaloshinskii Moriya exchange interactions and shown that this effect can be described by the double sine-Gordon model.
The investigation presented is sufficient to set up a complete picture of Lorentz TEM imaging of incommensurate magnetic structures of the model and trace its evolution with application of magnetic field and mechanical stress.
Through comparison with the dSG model we have identified the incommensurate 1s-phase by experimental Fresnel TEM imaging of a mechanically strained sample of the chiral helimagnet CrNb$_3$S$_6$, confirming a result first theoretically predicted by Iwabuchi \cite{Iwabuchi1983}.
Our results demonstrate an additional degree of freedom for control of the anisotropies present in chiral helimagnets for potential applications in the areas of spintronics and the emerging field of strain manipulated spintronics.

Original experimental data files are available at http://dx.doi.org/10.5525/gla.researchdata.1006.

\begin{acknowledgments}
This work was supported by a Grants-in-Aid for Scientific Research (B) (KAKENHI Grant Nos. 17H02767 and 17H02923) from the MEXT of the Japanese Government, JSPS Bilateral Joint Research Projects (JSPS-FBR), the JSPS Core-to-Core Program, A. Advanced Research Networks, the Engineering and Physical Sciences Research Council (EPSRC) of the U.K. under Grant No. EP/M024423/1, and Grants-in-Aid for Scientific Research on Innovative Areas `Quantum Liquid Crystals' (KAKENHI Grant No. JP19H05826) from JSPS of Japan.
I.P. and A.A.T. acknowledge financial support by the Ministry of Education and Science of the Russian Federation, Grant No. MK-1731.2018.2 and by the Russian Foundation for Basic Research (RFBR), Grant No. 18-32-00769 (mol\_a).
A.S.O. and A.A.T. acknowledge funding by the Foundation for the Advancement of Theoretical Physics and Mathematics BASIS Grant No. 17-11-107, and by Act 211 Government of the Russian Federation, contract No. 02.A03.21.0006. 
A.S.O. thanks the Russian Foundation for Basic Research (RFBR), Grant 20-52-50005, and the Ministry of Education and Science of Russia, Project No. FEUZ-2020-0054.
\end{acknowledgments}

\appendix

\section{Reduction of algebraic integrands to Jacobian elliptic functions}
From Eq.(\ref{DSG}) it is straightforward to see that 
\begin{equation} \label{1Int}
\left( \frac{d \varphi}{dz} \right)^2 =  b_1 \cos \varphi + b_2 \cos 2 \varphi + c,
\end{equation}
where $c$ is a constant of integration.  

The requirement of positiveness of the right hand side of this relation for $\varphi$ values where it reaches a minimum results in three different solution domains: 
$$
(\textrm{I}) \quad  v<0,  \quad u>1-v,
$$
$$
(\textrm{II})  \quad   0<v<\frac14,  \quad    1-v< u< v+ \frac{1}{8v},
$$
$$
(\textrm{III})  \quad   v>0,  \quad    u> v+ \frac{1}{8v},
$$
where $u=c/b_1>0$ and $v=b_2/b_1$. 

Integration of (\ref{1Int}) gives 
\begin{equation} \label{GenInt}
 z -z_0 = \mp \frac{1}{\sqrt{b_1}} \int^{x(z)}_{-1} \frac{dx}{\sqrt{(1-x)(1+x)(u+x+ 2vx^2-v)}},
\end{equation}
where $x=\cos \varphi$  and $\varphi(z_0)=\pi$.

In region (I), where $u+x+ 2vx^2-v = 2v(x-x_1)(x-x_2)$ with
\begin{equation} \label{x12}
x_{1,2} = \frac{-1 \pm \sqrt{1-8v(u-v)}}{4v},
\end{equation}
the hierarchy $x_1 < -1 < x(z) < 1 < x_2$ is fulfilled.

Then, the integration in Eq.(\ref{GenInt}) may be realized through the  inverse Jacobian elliptic function\cite{Byrd1971}, 
$$
\int^y_c \frac{dx}{\sqrt{(a-x)(b-x)(x-c)(x-d)}} = g\, \textrm{sn}^{-1}(\sin \varphi,k), 
$$
provided $a>b\geq y > c > d$. Here, 
$$
  g = \frac{2}{\sqrt{(a-c)(b-d)}},
$$
$$
\varphi = \sin^{-1} \left( \sqrt{\frac{(b-d)(y-c)}{(b-c)(y-d)}}  \right),
$$
and 
$$
k^2 = \frac{(b-c)(a-d)}{(a-c)(b-d)}.
$$
This ensures the solution (\ref{RegI}).

In region (II), the hierarchy 
$$
   x_2 < x_1 < -1 \leq x(z) \leq 1
$$
exists.  This enables us to use 
$$
\int^y_b \frac{dx}{\sqrt{(a-x)(x-b)(x-c)(x-d)}} = g \, \textrm{sn}^{-1}(\sin \varphi,k),
$$
where   
$$
\varphi = \sin^{-1} \left( \sqrt{\frac{(a-c)(y-b)}{(a-b)(y-c)}}  \right),
$$
and 
$$
k^2 = \frac{(a-b)(c-d)}{(a-c)(b-d)}
$$
provided $d<c<b<y \leq a$.  This results once more in Eq. (\ref{RegI}).

In the domain (III), one has $D=1-8v(u-v)<0$, therefore
$$
u+x+ 2vx^2-v = 2v \left[ \left( x+ \frac{1}{4v} \right)^2  +\frac{|D|}{16v^2}   \right].
$$
This allows us to apply 
$$
\int^y_b \frac{dt}{\sqrt{(a-t)(t-b)\left[ (t-b_1)^2 + a^2_1 \right]}} = g \, \textrm{cn}^{-1}(\cos \varphi, k)
$$
with   
$$
g = \frac{1}{\sqrt{AB}},
$$
$$
A = \sqrt{(a-b_1)^2+a^2_1}, \quad B = \sqrt{(b-b_1)^2+a^2_1},  
$$
$$
k^2 = \frac{(a-b)^2-(A-B)^2}{4AB},
$$
$$
\varphi = \cos^{-1} \left[ \frac{(a-y)B-(y-b)A}{(a-y)B+(y-b)A} \right]
$$
provided $a \geq y > b$. The integration yields (\ref{RegIII}).

\section{The thermodynamical potential}
The energy per unit length, measured in units $J_{||}M^2_s/2$,   
$$
\mathcal{E} =  \frac{1}{L_0}    \int^{L_0}_0 dz \, \left[ 
 \left( \frac{d \varphi}{dz} \right)^2 - 2 \frac{D}{J_{||}} \left( \frac{d \varphi}{dz} \right) 
+  b_1 \cos \varphi + b_2 \cos 2 \varphi
\right] 
$$
may be reduced to the form 
\begin{equation}  \label{ExprForE}
\mathcal{E} = - c - \frac{4\pi q_0}{L_0} +   \frac{2}{L_0}  \int^{L_0}_0 dz  \left( \frac{d \varphi}{dz} \right)^2, 
\end{equation}
where the result (\ref{1Int}) is used.

{\it 1s-phase.} By substituting here the solution  (\ref{RegI}) and carrying out integration with respect to the coordinate $z$, we get
\begin{widetext}
\begin{equation} \label{EnS1}
\mathcal{E}  = - c - \frac{\pi q_0}{K} \sqrt{2b_2(x_2+1)(x_1-1)}  + 2b_2 (x_2+1)(x_1-1) \frac{E}{K} - 2b_2 (x^2_1-1) +2b_2 (x_1+1)(x_2+x_1) \frac{\Pi}{K},
\end{equation}
where $\Pi$ is the complete elliptic integral of the third kind 
$$
\Pi \left( \alpha^2, \kappa \right) = \int^K_0 \frac{du}{1-\alpha^2 \textrm{sn}^2 u} 
$$
with $\alpha^2 =2/(1-x_1)$ and $0 < \alpha^2 < \kappa^2 <1$.  To obtain (\ref{EnS1}) we  make use of the result
$$
 \int^K_0 \frac{du}{\left( 1-\alpha^2 \textrm{sn}^2 u \right)^2} = \frac{1}{2(1-\alpha^2)(\kappa^2 - \alpha^2)}
\left[
-\alpha^2 E + \left( \alpha^4 - 2\alpha^2 \kappa^2 -2 \alpha^2 + 3 \kappa^2 \right) \Pi + \left( \alpha^2 - \kappa^2 \right) K
\right]. 
$$

{\it 2s-phase.}  Making an evaluation of $\varphi^{'}_z$ with the aid of the solution (\ref{RegIII}), we obtain as the final result
\begin{equation} \label{EnS2}
\mathcal{E}  = b_2 - \frac{b_1 l}{2} + \frac{b_1}{2l} - \frac{b^2_1}{4 b_2}  - \frac{\pi q_0}{K}  \sqrt{\frac{b_1 (l^2-1)}{2l}}
- b_1 \frac{(1-l^2)}{l} \frac{E}{K} + b_1 \left( l + \frac{1}{4v} \right) \frac{\Pi}{K}.
\end{equation}
 \end{widetext}
This expression involves the elliptic integrals for which $\alpha^2 = -4 l v$ should be chosen.
In contrast with the 1s-phase, there is the hierarchy $0<\kappa^2<\alpha^2<1$.
To eliminate the parameter $c$ in Eq.(\ref{ExprForE}), we made use the relationship $u = -v -l/2 - 1/(2l)$.

\section{Fourier transform of the 1s-phase}
Most notably, we note the result
\begin{equation} \label{sinphi}
\sin \varphi = \pm \dfrac{2\sqrt{x^2_1-1}\, \textrm{sn}(\bar{z}) \textrm{cn}(\bar{z})}{2\textrm{sn}^2(\bar{z})-1 +x_1}
\end{equation}
originated from Eq.(\ref{RegI}).

In determining the parameter $b$ such that 
\begin{equation} \label{bRI}
 \textrm{sn}^2( b,\kappa) = \frac{x_2+1}{x_2-x_1},
\end{equation}
we have $0< \textrm{sn}^2 b<1$ in region (I), where $x_1<-1$ and $x_2>1$.

Taking account of Eq. (\ref{k2phase1}), one may present (\ref{sinphi}) in the form
$$
 \sin \varphi  =  \frac12 \sqrt{(x_2+1)(1-x_1)} \left[ \textrm{dn}(\bar{z}+b,\kappa) - \textrm{dn}(\bar{z}-b,\kappa)    \right],
$$
where the positive sign is taken for certainty.  

A required result is achieved through the series expansion
\begin{equation} \label{FSdn}
\textrm{dn} (u) = \frac{\pi}{2K} + \dfrac{\pi}{2K} \sum_{n\not=0} \frac{\exp\left( in\dfrac{\pi}{K}u \right)}{\cosh \left( \pi \dfrac{K'}{K} n \right)}. 
\end{equation}

Calculation of the magnetic phase shift gives rise to 
$$
\phi(y,z) = - \frac{4e\mu_0 M_0 d}{\hbar \sqrt{-2b_2}}  \sum_{n=1}^{\infty} \dfrac{\cos \left( \dfrac{\pi n \bar{z}}{K} \right) \sin  \left( \dfrac{\pi n b}{K} \right)}{n \cosh \left( \dfrac{\pi n K'}{K} \right)}
$$
\begin{equation}
= - \frac{2e\mu_0 M_0 d}{\hbar \sqrt{-2b_2}}  \left[
  \textrm{am} (\bar{z}+b) - \textrm{am} (\bar{z}-b) - \dfrac{\pi}{K}b
\right]
\end{equation}
that amounts to (\ref{PhaseShift1}), keeping in mind the Fourier series for the Jacobi's amplitude function
$$
\textrm{am} (u) = \frac{\pi u}{2K} +  \sum_{n=1}^{\infty} \frac{\sin \left( n \frac{\pi}{K} u \right)}{n \cosh \left( \pi \dfrac{K'}{K} n \right)}.
$$

Similarly, the parameter $b$ may be introduced for the II region, where $x_2<x_1<-1$, such  that 
\begin{equation} \label{bRII}
  \textrm{sn}^2 (b,\kappa') = \frac{x_2+1}{x_2-1},
\end{equation} 
with $\kappa' = \sqrt{1-\kappa^2}$ is being the complementary modulus.

As a next step, consider the following expression 
\begin{equation} \label{scdif}
\frac12 \sqrt{(x_1+1)(x_2-1)} \left[ \textrm{sc}(\bar{z}-ib,\kappa) + \textrm{sc}(\bar{z}+ib,\kappa)  \right]. 
\end{equation} 
By using addition theorems for the Jacobi elliptic functions $\textrm{cn}(\ldots)$ and $\textrm{sn}(\ldots)$  it may be reduced to 
$$
\sqrt{(x_1+1)(x_2-1)} \textrm{sn}(\bar{z}) \textrm{cn}(\bar{z})  \textrm{dn}(i b)
$$ 
$$
\times \frac{\textrm{cn}^2(i b)+\textrm{sn}^2(i b)  \textrm{dn}^2(\bar{z})}{\textrm{cn}^2(\bar{z})\textrm{cn}^2(i b)-\textrm{sn}^2(\bar{z})\textrm{sn}^2(i b) \textrm{dn}^2(\bar{z}) \textrm{dn}^2(\bar{ib})}.
$$
This result may be simplified via the Jacobi imaginary transformations 
$$
\textrm{sn}^2(i b,\kappa) = - \frac{\textrm{sn}^2(b,\kappa')}{\textrm{cn}^2(b,\kappa')} =  \frac12 (1+x_2), 
$$
$$
\textrm{cn}^2(i b,\kappa) = \frac{1}{\textrm{cn}^2(b,\kappa')} = \frac12 (1-x_2),
$$
$$
\textrm{dn}^2(i b,\kappa) =  \frac{\textrm{dn}^2(b,\kappa')}{\textrm{cn}^2(b,\kappa')} =  \frac{1-x_2}{1-x_1}
$$
which yields
\begin{equation} \label{sin1}
\pm \frac{2  \sqrt{x^2_1-1} \, \textrm{sn}(\bar{z}) \textrm{cn}(\bar{z}) }{1-x_1-2\textrm{sn}^2(\bar{z})}.
\end{equation}
This outcome is nothing but $\sin \varphi$ as evident in Eq.(\ref{sinphi}).

Going back to Eq.(\ref{scdif})  it can be seen that periodicity of elliptic functions transforms this expression into 
\begin{equation}  \label{sin2}
 - \frac{i}{2} \sqrt{(x_1-1)(x_2+1)} 
\end{equation}
$$
\times \left[ \textrm{dn}(\bar{z}-K + i(K'-b),\kappa) - \textrm{dn}(\bar{z}-K - i(K'-b),\kappa)  \right]. 
$$
By combining Eqs.(\ref{sin1}, \ref{sin2}) we obtain 
\begin{equation}
\sin \varphi =  - \frac{i}{2} \sqrt{(x_1-1)(x_2+1)} 
\end{equation}
$$
\times   \left[ \textrm{dn}(\bar{z}-K + i(K'-b),\kappa) - \textrm{dn}(\bar{z}-K - i(K'-b),\kappa)  \right]. 
$$ 

Once again using the Fourier series expansion (\ref{FSdn}) we get
$$
\phi(y,z) = - \frac{4e\mu_0 M_0 d}{\hbar \sqrt{2b_2}}  \sum_{n=1}^{\infty} \dfrac{ \sinh  \left( \dfrac{\pi n}{K} \left[ K' - b \right] \right)}{n \cosh \left( \dfrac{\pi n K'}{K} \right)}  
$$
$$
\times \cos \left( \dfrac{\pi n }{K} \left[ \bar{z} -K \right] \right)
$$
which provides the answer (\ref{Phi2}).

\section{Fourier transform of the 2s-phase}
In advance, we note that
$$
\sin \varphi = \pm \frac{\sqrt{1-l^2} \, \textrm{sn}(\bar{z})}{1-l  \textrm{cn}(\bar{z})}
$$
stemming from Eq.(\ref{RegIII})  and introduce the parameter $b$, $0<b<K'$, that obeys the equations
$$
 \textrm{sn}^2( b,\kappa') = 1- l^2, \quad 
\textrm{cn}^2( b,\kappa') = l^2,
$$
\begin{equation}  \label{bParam}
\textrm{dn}^2( b,\kappa') = -4lv. 
\end{equation}

As an alternative, one may verify the result 
$$
\sin \varphi = \frac{\sqrt{1-l^2}}{2  \textrm{dn}( b,\kappa')}  \left[
   \textrm{ns}(\bar{z}-i b,\kappa) + \textrm{ns}(\bar{z}+ i b,\kappa) 
\right.
$$
$$
\left.
   -  \textrm{cs}(\bar{z}- i b,\kappa) -  \textrm{cs}(\bar{z}+ i b,\kappa)
\right]
$$
which can be easily proved by means of the definition (\ref{kappa2s}), the  Jacobi imaginary transformations 
$$
 \textrm{dn}( ib,\kappa) =  \textrm{dc}( b,\kappa') = - i \sqrt{4v/l}, 
$$
$$
 \textrm{sn}( ib,\kappa) = i  \textrm{sc}( b,\kappa') = - i \sqrt{1-l^2}/{l},  
$$
$$
\textrm{cn}( ib,\kappa) = \textrm{nc}( b,\kappa') = - {1}/{l}
$$
and the addition theorems for the Jacobi elliptic functions. 

Using periodicty of the elliptic functions
$$
 \textrm{ns}(\bar{z} \pm i b,\kappa) = \kappa  \textrm{sn}\left[\bar{z} \mp  i(K'- b),\kappa \right],
$$
$$
\textrm{cn}(\bar{z} \pm  i b,\kappa) = \mp \frac{i}{\kappa} \textrm{ds}\left[\bar{z} \mp  i(K'- b),\kappa \right],
$$
one may reach the endpoint convenient  for a Fourier series expansion
$$
\sin \varphi = \frac{\kappa \sqrt{1-l^2}}{2  \textrm{dn}( b,\kappa')}   \left\{
     \textrm{sn}\left[\bar{z} +  i(K'- b),\kappa \right]  + \textrm{sn}\left[\bar{z} -  i(K'- b),\kappa \right]
\right.
$$
\begin{equation}
\left.
+ \frac{i}{\kappa} \left(
   \textrm{dn}\left[\bar{z} -  i(K'- b),\kappa \right]
- \textrm{dn}\left[\bar{z} +  i(K'- b),\kappa \right]
\right)
\right\},
\end{equation}
where
\begin{equation} \label{FSsn}
\textrm{sn} (u) =  \dfrac{\pi}{2iK\kappa} \sum_{n=-\infty}^{\infty}  \dfrac{\exp \left[ \dfrac{i\pi u}{2K} (2n+1) \right]}{\sinh \left[ \dfrac{\pi K'}{2K} (2n+1) \right]}
\end{equation}
and $\textrm{dn} (\ldots)$ is given by Eq.(\ref{FSdn}).

Then the corresponding result for the magnetic phase shift
$$
\phi(y,z) =  2 \frac{e\mu_0 M_0 d}{\hbar \sqrt{2b_2}}
$$
$$
\times \left\{
   \sum_{n=0}^{\infty} \frac{\cos \left[ \frac{\pi}{2K}(2n+1) \bar{z} \right] \cosh \left[ \frac{\pi}{2K}(2n+1) (K'-b) \right]}{(n+1/2) \sinh \left[ \frac{\pi K'}{2K}(2n+1) \right]}
\right. 
$$
$$
\left.
-  \sum_{n=1}^{\infty} \frac{\cos \left[ \frac{\pi}{K}n \bar{z} \right] \sinh \left[ \frac{\pi}{K}n (K'-b) \right]}{n \cosh \left[ \frac{\pi K'}{K}n \right]}
\right\}
$$
may be simplified to the form (\ref{PhaseShift2s}),  where the relationship 
$$
\int^u_K dt \, \textrm{sn} (t) = \frac{1}{2\kappa} \ln \left( \dfrac{1-\kappa \textrm{cd} u}{1+\kappa \textrm{cd} u} \right) 
$$
should be accounted for.

\onecolumngrid


\begin{thebibliography}{99}
\bibitem{OurComment} In Ref. [45] it was shown that $\kappa^2=0$ is not a singular point of the thermodynamic potential subjected to the transformation $\kappa^2 \to - \kappa^2$, since the zero value is not singular for elliptic functions and integrals. We emphasize also that the 1$\bar{\textrm{s}}$ and 2s phases have equal periods, $4K \left(q^2_0 - 2b_2\right)^{-\frac12}$, at the OC boundary.
\end{thebibliography}
\end{document}